\documentclass[a4paper,11pt]{article}
\pdfoutput=1

\usepackage{float}
\usepackage{jheppub}
\usepackage{amsmath,amsfonts,amssymb,amsthm,graphicx,color}
\usepackage{mathrsfs}
\usepackage[dvipsnames]{xcolor}
\usepackage{verbatim}
\usepackage{enumerate}
\usepackage{epsfig}
\usepackage{multirow}
\usepackage{comment}
\usepackage{datetime}
\usepackage{slashed}
\usepackage{framed}
\usepackage{longtable}
\usepackage[mathscr]{euscript}
\usepackage{cancel}
\usepackage{tensor}
\usepackage{mathtools}
\usepackage{caption}
\usepackage{subcaption}
\usepackage[multiple]{footmisc}
\usepackage{pgfplots}
\usepackage[T1]{fontenc}
\usepackage[utf8]{inputenc}
\usepackage{tikz}
\usepackage{placeins}
\usepackage{hyperref}
\usetikzlibrary{decorations.markings}
\usepackage{multirow}

\newcommand{\nn}{\nonumber}

\def\be{\begin{equation}}
\def\ee{\end{equation}}
\def\bea{\begin{align}}
\def\eea{\end{align}}
\def\beaq{\begin{eqnarray}}
\def\eeaq{\end{eqnarray}}

\def\a{\alpha} \def\b{\beta} \def\g{\gamma} 
   \def\q{\theta}
 \def\k{\kappa}  \def\m{\mu}
\def\n{\nu}    
\def\s{\sigma} \def\t{\tau}

   \def\L{\Lambda}

\title{Charge (in)stability and superradiance of Topological Stars}

\author{Andrea Cipriani,}
\author{Carlo Di Benedetto,}
\author{Giorgio Di Russo\footnote{Address after February 9th: Institut de Physique Th\'eorique, Universit\'e Paris-Saclay, CEA, Orme des
Merisiers, Gif-sur-Yvette, 91191 CEDEX, France},}
\author{Alfredo Grillo,}
\author{Giuseppe Sudano}

\affiliation{Dipartimento di Fisica, Universit\`a di Roma ``Tor Vergata" \& Sezione INFN Roma2, Via della Ricerca Scientifica 1, 00133, Roma, Italy}

\abstract{We study linear massive scalar charged perturbations of Topological Stars in the fuzzball and in the black hole (Black String) regimes. The objects that naturally couple to the electric 3-form field strength of these solutions are charged strings, wound around the compact direction. We explore the possibility of instabilities of these solutions, in analogy with the charge instability already highlighted for other non-BPS geometries like JMaRT. This issue is addressed by calculating quasi-normal mode frequencies with a variety of techniques: WKB approximation, direct integration, Leaver method and by exploiting the recently discovered correspondence between black hole/fuzzball perturbation theory and quantum Seiberg-Witten curves. All mode frequencies we find have negative imaginary parts, implying an exponential decay in time. This suggests a linear stability of Topological Stars also in this new scenario. In addition, we study the charge superradiance for the Black String. We compute the amplification factor with the numerical integration method and a quantum Seiberg-Witten motivated definition including instantonic corrections.}

\begin{document}
\maketitle
\flushbottom 
\section{Introduction}
Black holes (BH) are nowadays regarded as the ideal playground to focus on when attempting to solve the lasting clash between Quantum Mechanics and General Relativity. As a matter of fact, a complete Quantum Gravity theory should be able to puzzle out some inconsistencies regarding BH thermodynamical properties, e. g. the microscopic origin of their entropy and the information paradox \cite{Mathur:2009hf, Bena:2022rna}. One of the major successes of String Theory has been the counting of microstates for a supersymmetric BH, in the limit of vanishing gravitational coupling \cite{Sen:1995in, Strominger:1996sh}. On the other hand, the resolution of the information paradox also requires to know the structure of these microstates at finite coupling. In this respect, one suggestion is represented by the `fuzzball proposal' \cite{Lunin:2001fv, Mathur:2005zp}: singular and horizon-full BHs should be thought as the superposition of quantum states, some of which admit a classical description as smooth and horizonless gravity solutions with the same conserved quantities (mass, electric charge and angular momentum) of the BH. Remarkably, these geometries emit radiation preserving quantum information. 

The identification of some (a)typical microstate geometries has been rather successful for supersymmetric BHs \cite{Giusto:2004id, Giusto:2004ip, Giusto:2011fy, Turton:2012ny, Bena:2015bea, Bena:2016agb, Bena:2016ypk, Bena:2017xbt, Bena:2017upb, Bianchi:2016bgx, Bianchi:2017bxl}. This task has been more challenging for non-BPS or even non-extremal BHs (which is the ultimate goal if one aims to find microstate geometries for astrophysical BHs). For instance, JMaRT solitons \cite{Jejjala:2005yu}, the first known non-BPS smooth horizonless geometries, are not viable microstate geometries: a BH with the same charges would be over-rotating, thus its horizon would collapse leading to a naked singularity. Moreover, due to the presence of an ergo-region without horizon, JMaRT solitons are both affected by an ergo-region instability \cite{Cardoso:2005gj} and a charge instability \cite{Bianchi:2023rlt}. Very recent developments in the fuzzball programme for non-BPS geometries also include microstrata \cite{Ganchev:2021pgs, Ganchev:2023sth} and Schwarzschild-like topological solitons \cite{Bah:2022yji}. 

In this paper, we focus on a recently discovered smooth horizonless non-BPS geometry called Topological Star (TS) \cite{Bah:2020ogh, Bah:2020pdz, Heidmann:2021cms, Bah:2021owp, Heidmann:2022zyd}. This is a static solution of Einstein-Maxwell theory in 5 dimensions, whose 4-dimensional reduction is a BH solution of Einstein-Maxwell-Dilaton theory. A TS is characterized by a non-trivial topology and its Gregory-Laflamme instability \cite{Gregory:1993vy} is avoided by turning on electromagnetic fluxes. Specifically, the magnetic field of a TS is sourced by a magnetic monopole, whereas the electric field is sourced by a string wound around the compact direction. A double Wick rotation or an analytical continuation of the parameters \cite{Bah:2021irr} maps the TS to a black object with the same mass and charges. This is called Black String (BS) and is a two-charge non-rotating instance of a BH in the Cvetic-Youm class \cite{Cvetic:1996xz, Cvetic:1997uw}.

The study of BHs is not only appealing for theoretical reasons: since a few years ago, new observational techniques have been yielding outstanding results. LIGO and Virgo interferometers \cite{LIGOScientific:2016aoc, LIGOScientific:2016lio} have been able to detect a signal of gravitational waves coming from the late phase (ring-down) of the merger of two BHs. Moreover, the Event Horizon Telescope \cite{EventHorizonTelescope:2019dse} has provided the first direct images of the light ring and the accretion disk of a BH. On grounds of these outcomes, one might hope to directly test novel ideas in BH physics or even Quantum Gravity, through direct observations. In particular, these may allow to discriminate between BHs and fuzzball geometries due to their different dynamical properties. In addition, the fuzzball conjecture could be tested as opposed to alternative models \cite{Cardoso:2016rao, Cardoso:2017cqb, Cardoso:2019rvt}, collectively called Exotic Compact Objects (ECO). A non-exhaustive list of the latter also includes boson stars \cite{Schunck:2003kk}, gravastars \cite{Mazur:2001fv, Danielsson:2017riq}, firewalls \cite{Almheiri:2012rt} and wormholes \cite{Solodukhin:2005qy} (also see \cite{Balasubramanian:2022gmo, Balasubramanian:2022lnw, Climent:2024trz} for recent progress). 

The dynamical aspects of gravitational objects can be examined by probing them with particles, waves or strings \cite{Bianchi:2017sds}. BHs and fuzzballs can be discriminated by identifying observables with different behaviours for the two types of geometries: the shape of the light rings \cite{Bianchi:2018kzy, Bianchi:2020des, Bianchi:2020yzr}, the spectrum of Quasi-Normal Modes (QNMs) \cite{Chowdhury:2007jx, Chakrabarty:2019ujg, Bena:2020yii, Ikeda:2021uvc}, the Tidal Love Number (TLN) \cite{Consoli:2022eey, Bianchi:2023sfs, DiRusso:2024hmd} and the multi-polar structure \cite{Bena:2020see, Bianchi:2020bxa, Bena:2020uup, Bianchi:2020miz, Mayerson:2020tpn, Bah:2021jno}. Superradiance \cite{Brito:2015oca} is another significant phenomenon in this context. It consists in the amplification of wave perturbations scattering off BHs, which thereby lose energy, charge and angular momentum. Superradiance has been introduced and mostly studied \cite{Press:1972zz, Teukolsky:1974yv} for rotating BHs, where it can be considered as the wave counterpart of the Penrose process; nonetheless it has been also highlighted to occur for static charged BHs \cite{DiMenza:2014vpa, Benone:2015bst, Baake:2016oku}. Remarkably, superradiance can only take place for black objects and not for fuzzball geometries which, instead, preserve information. 

The purpose of the present investigation is to explore the dynamical behaviour of TS and BS under charged linear scalar perturbations. A single TS is not a good candidate as a microstate geometry for astrophysical BHs \footnote{Indeed, TSs usually constitute the building blocks of more complicated multi-bubble solutions \cite{Bah:2021owp}.} (e.g., it is magnetically charged and it has been proven to be a `spherical spacetime mirror' for photons \cite{Heidmann:2022ehn}). Nevertheless, because of the large symmetry of the TS metric, the dynamics of particles or waves in their background is separable. Relevant information about their dynamical features can be easily extracted: TSs, then, represent very useful tools to learn or to test the typical behaviour expected for more general smooth horizonless geometries. 

The response of TS under linear scalar massive or massless perturbations has been already studied in \cite{Heidmann:2022ehn, Heidmann:2023ojf, Bianchi:2023sfs}. In particular, photon geodesics, QNMs and TLNs in the TS background have been analyzed, allowing to argue for the linear stability of TSs. Moreover, being their dynamical TLN real, TSs have been proved to undergo deformations only and not dissipation, consistently with the absence of a horizon. 

In the aforementioned works, probes have been chosen to be chargeless or at most to have Kaluza-Klein momentum along the compact direction. In the present paper, instead, we investigate the possibility of instabilities when the probes are \textit{charged} under the electric field of a TS, in analogy with the charge instability pointed out for JMaRT geometries \cite{Bianchi:2023rlt}. The electric gauge field of a TS is a 2-form, sourced by an electrically charged string, wound around the compact direction. Differently from the previous treatments, probes are not point-like particles, but closed strings wound around the compact direction. These could be identified as some of the winding modes of TS's electric string, after they slide off the bolt during the discharge mechanism. 

We will make use of the correspondence between BH/fuzzball perturbation theory and quantum Seiberg - Witten (qSW) curves of $\mathcal{N} = 2$ SYM theory with gauge group SU(2) \cite{Aminov:2020yma}. Besides providing a further tool to compute QNMs \cite{Bianchi:2021xpr, Bianchi:2021mft, Bianchi:2022qph, Bianchi:2023rlt, Bianchi:2023sfs} this correspondence has allowed to devise a definition for the dynamical TLN \cite{Consoli:2022eey, Bianchi:2023sfs, DiRusso:2024hmd}, by taking into account instantonic corrections. In all these developments, a prominent role has been played by the Heun connection formulae \cite{Bonelli:2021uvf, Bonelli:2022ten}, recently found exploiting the Alday-Gaiotto-Tachikawa (AGT) duality \cite{Alday:2009aq}. In particular, the radial equation for waves in TS/BS background is a Confluent Heun Equation (CHE), which can be mapped to the CHE arising from a qSW curve with three massive hypermultiplets. Alternative semi-analytical procedures for the computation of QNMs are based on integrability techniques \cite{Fioravanti:2021dce, Fioravanti:2020udo, Rossi:2022mwx}. 

From the QNM spectrum, we infer that TSs are stable also when probed with scalar \textit{charged} perturbations. Indeed, only modes with negative imaginary part have been found. We cannot rule out instabilities under vector or tensor perturbations as well as non-linear instabilities \cite{Eperon:2016cdd} also discovered for BPS geometries \cite{Giusto:2004id, Giusto:2004ip}. Some further insights will be reported in a future work \cite{DiRusso:2024xy}. 

Being static geometries, TSs and BSs cannot give rise to rotation superradiance. We investigate, instead, the possibility of charge superradiance. No amplitude increases have been discovered for TSs at any frequencies. This is consistent with the fact that information is preserved by fuzzball geometries. By contrast, BSs can be well-suited to examine charge superradiance. We show the presence of an enhancement of wave amplitudes in a certain range of frequencies for BSs, through the analysis of the amplification factor. On grounds of the BH/fuzzball perturbation theory - qSW curves correspondence, we have added the instantonic corrections to the definition of this observable. The results have been compared with a corresponding numerical integration. The absence of mirror mechanisms, though, excludes the possibility of superradiant instabilities \cite{Press:1972zz}.

This paper is organised as follows. In Section \ref{Top star} we introduce the TS solution and discuss the coupling of its electric field with electrically charged strings. We write the wave equation for charged linear scalar perturbations and separate the radial and the angular dynamics. In Section \ref{quantumSW} we review the theory of quantum Seiberg-Witten curves for $\mathcal{N} = 2$ super Yang-Mills with gauge group $SU(2)$; we perform the matching with the perturbation theory in TS background and define the quantization condition for the computation of QNMs. Section \ref{Leaver} illustrates Leaver method to compute QNMs for the TS. The frequencies of the charged QNMs of a TS calculated with the latter methods are contained in Section \ref{TopstarQNM} and compared with those arising from a numerical integration. Section \ref{Charge superradiance} investigates charge superradiance for the BS. We compute the near-extremal near-superradiant QNMs and the prompt ring-down modes. The employed techniques are: WKB approximation, numerical integration, Leaver method and the correspondence with qSW curves. Finally, the amplification factor is defined and corrected with the instantonic contributions. The outcome is compared with a numerical prediction. In Section \ref{Conclusion} we draw our conclusions and suggest possible outlooks. In Appendices we report further details. Appendix \ref{Charged wire} is focused on the electric field generated by an electrically charged wire in a spacetime of generic dimension. In Appendix \ref{appmeas} we discuss the measure units and the dimensions of the observables we study. Appendix \ref{appA} describes the numerical method used to compute the amplification factor.

\section{Charged scalar perturbations in a Topological Star background} \label{Top star}
In this Section we present the solution of a TS. We probe this geometry with charged strings with respect to its electric field. Accordingly, we devise a shift of variables in the scalar wave equation, that takes into account the stringy nature and the charge of the probes.

\subsection{The background}
TSs are smooth horizonless static solutions of Einstein-Maxwell theory in 5 dimensions, described by the action \cite{Bah:2020ogh, Bah:2020pdz}
\be
S_5=\int \mathrm{d}^5x \, \sqrt{-\det{g}}\left({\frac{R}{2\k_5^2}}-{\frac{1}{2}}\bigl|F^{(m)}\bigr|^2-{\frac{1}{2}}\bigl|F^{(e)}\bigr|^2\right)
\ee
The corresponding equations of motion are\footnote{In this paper, $\varepsilon_{01234} = +1$.}
\be \label{eqmotion}
\mathrm{d} \star F^{(m)}= \mathrm{d} \star F^{(e)}=0\quad,\quad R_{\mu\nu}=\k_5^2\left(T_{\mu\nu}-{\frac{1}{3}}g_{\mu \nu}{T_\a}^\a\right)
\ee
where $T_{\mu\nu}$ is the stress energy tensor
\be
T_{\mu\nu}={F^{(m)}}_{\m\a}{{F^{(m)}}_\nu}^\a-{\frac{1}{4}}g_{\mu\nu}{F^{(m)}}_{\a\b}{F^{(m)}}^{\a\b}+{\frac{1}{2}}\Bigg[{F^{(e)}}_{\mu\a\b}{{F^{(e)}}_\n}^{\a\b}-{\frac{1}{6}}g_{\mu\nu}{F^{(e)}}_{\a\b\g}{F^{(e)}}^{\a\b\g}\Bigg]
\ee
The metric of a TS is given by
\be \label{eq:top}
\mathrm{d}s^2=- f_s(r) \, \mathrm{d}t^2 + {\frac{\mathrm{d}r^2}{f_b (r) f_s (r)}} + r^2 \,(\mathrm{d}\q^2 + \sin^2\q \, \, \mathrm{d} \phi^2) + f_b(r) \, \mathrm{d}y^2
\ee
where
\be \label{eq:warping}
f_s(r) = 1-\frac{r_s}{r}  , \quad f_b(r) = 1-\frac{r_b}{r}
\ee
In \eqref{eq:top}, $t$ and $(r, \q, \phi)$ are the time and the spherical coordinates of the 4-dimensional spacetime, while $y$ is a further compact coordinate such that $y \sim y + 2 \pi R_y$. The metric of a TS is then static, asymptotically $\mathbb{R}^{1, 3} \times S^1_y$ and spherically symmetric in 4 dimensions\footnote{This gravity solution can be seen as the harmonic superposition of a Schwarzschild String (4-dimensional Schwarzschild BH $\times S^1_y$) and a static Bubble of Nothing (an Euclidean Schwarzschild solution with a time direction). Introduced to describe an instability of the Kaluza-Klein vacuum \cite{Witten:1981gj}, the Bubble of Nothing has been recently studied to understand the stability of non-SUSY vacua \cite{Ooguri:2017njy,  GarciaEtxebarria:2020xsr, Dibitetto:2020csn, Bomans:2021ara} and in the framework of the Cobordism Conjecture \cite{Delgado:2023uqk}.}.

Three singularities can be identified for \eqref{eq:top}: 
\begin{itemize}
\item $r=0$ is a full-fledged curvature singularity; 
\item $r =r_s$ is a horizon, being such that the component $g_{tt}$ of the metric changes sign; 
\item $r = r_b$ is the bolt singularity, where the compact $y$-circle degenerates to a point. 
\end{itemize}
In correspondence of the latter singularity, the spacetime smoothly terminates at a cap. According to the location of the singularities we can distinguish two kinds of solutions:
\begin{itemize}
\item TS for $r_b > r_s > 0$: the horizon and the singularity are pushed out of the geometry, which is then smooth and horizonless; 
\item BS for $r_s \geq r_b>0$: the spacetime is horizon-full. 
\end{itemize}
The two different regimes are linked by the double analytical continuation or exchange of parameters
$$
(t, y) \leftrightarrow (i t, i y) \quad \mathrm{or} \quad r_s \leftrightarrow r_b
$$

The presence of electromagnetic fluxes prevents the gravitational collapse of the solution. The electric and magnetic field strengths are respectively given by 
\be \label{eq:fieldstrength}
F^{(e)}=\frac{Q}{r^2} \, \mathrm{d}r \wedge \mathrm{d}t\wedge \mathrm{d}y\quad,\quad F^{(m)}=P\sin\q \, \mathrm{d} \q \wedge \mathrm{d}\phi
\ee
The equations of motion \eqref{eqmotion} are solved if:
\be \label{eq:TopCharge}
P^2+Q^2=\frac{3 r_b r_s}{2\k_5^2}
\ee
The source for the field strength $F^{(e)}$ is a charged line wound along the compact direction $y$. Indeed this is the object that naturally couples to the 2-form gauge potential:
\be \label{eq:gaugepotential}
B^{(e)} = -\frac{Q}{r} \, \mathrm{d} t \wedge \mathrm{d} y
\ee
In order to understand the nature of the parameter $Q$ in \eqref{eq:fieldstrength} and \eqref{eq:gaugepotential}, in Appendix \ref{Charged wire} a reference to what happens to the case of an infinitely extended and uniformly charged wire is discussed. The final result is that $Q$ can be interpreted as the linear charge density of this wire and this implies, in particular units discussed in the Appendix \ref{appmeas}, that $[Q] = [L]$, consistently with \eqref{eq:TopCharge}. 

The absence of Gregory-Laflamme \cite{Gregory:1993vy} and thermodynamical instabilities for the TS further requires that \cite{Bah:2021irr} (the analogous condition for the BS arises after $r_b \leftrightarrow r_s$)
\be
r_s < r_b < 2 r_s
\ee
Finally, \eqref{eq:top} admits a 4-dimensional reduction to the Einstein-Maxwell-Dilaton theory. This allows to deduce the corresponding ADM mass
\be
\label{ADM mass}
\mathcal{M} = \frac{2 \pi}{\kappa_4^2} (2 r_s + r_b)
\ee
This expression holds both for the TS and for the BS. Remarkably, TSs are non-BPS fuzzball geometries, since extremality is obtained for $r_b = r_s$ (hence just for one particular instance of BS).

\subsection{Stringy probes in the Topological Star background} \label{Stringy}
We want to investigate the response of TS background to charged probes. Both the source and the probes of the gauge potential \eqref{eq:gaugepotential} are strings wound along the compact direction\footnote{Stringy probes have been also helpful to explore the phenomenon of tidal scrambling in fuzzball geometries \cite{Martinec:2020cml, Ceplak:2021kgl}.}. We expect the probe strings to originate from the discharge mechanism of the background. In particular, TS geometry smoothly caps off at $r = r_b$ hence the winding modes of the source can slip off the bolt singularity with no obstacles. However, we work in the probe limit.

Let us focus on the instance of rigid probes, which namely propagate without oscillations. Under this assumption, the embedding coordinates of their worldsheet in the TS geometry are 
\be \label{eq:embedding}
X^M (\tau, \sigma) = (X^\mu (\tau), \, Y = y^0 + n \sigma R_y) \quad, \quad \sigma \in [0, 2\pi]  
\ee
The spacetime indices are split as $M = (\mu, y)$, with $\mu = (t,r,\theta,\phi)$. ($\tau$, $\sigma$) are the (dimensionless) worldsheet coordinates of the string, $n$ is an integer and corresponds to the winding number.

The dynamics of the string in TS background is governed by the non-linear sigma model \cite{Callan:1985ia}
\be \label{eq:sigmamodel}
S  = \frac{1}{4 \pi \alpha'} \int_\Sigma \mathrm{d}\sigma \mathrm{d}\tau \sqrt{-h} \Big [ G_{MN} (X) \partial_a X^M \partial_b X^N h^{ab} + B_{MN} (X) \partial_a X^M \partial_b X^N \varepsilon^{ab} \Big ]
\ee
In this action, $\Sigma$ represents the worldsheet of the string, $h^{ab} = \mathrm{diag}(-+)$ is the worldsheet metric, $\varepsilon_{ab}$ is the 2-dimensional antisymmetric tensor (such that $\varepsilon_{\t \s} = 1$). Moreover $G_{MN}$ and $B_{MN}$ are the metric \eqref{eq:top} and the gauge potential \eqref{eq:gaugepotential} of the background (the dilaton is missing in the TS solution). 

The conjugate momenta to the spacetime coordinates in \eqref{eq:sigmamodel} read 
\be \label{eq:momentum}
P_M = \frac{\partial L}{\partial \dot{X}^M}
\ee
where $\dot{X}^M = \partial_\t X^M$ and the Lagrangian $L = L\bigl [X^M, \partial_a X^M \bigr]$ is such that 
$$
S = \int \mathrm{d} \t \, L
$$
The effect of the gauge field on the expression of the conjugate momenta can be studied by splitting the Lagrangian as 
\be
L = L_G + L_B
\ee
with 
$$
L_G = \frac{1}{4 \pi \alpha'} \int_0^{2 \pi} \mathrm{d}\sigma \sqrt{-h} \, G_{MN} (X) \partial_a X^M \partial_b X^N h^{ab}
$$
$$
L_B = \frac{1}{4 \pi \alpha'} \int_0^{2 \pi} \mathrm{d}\sigma \sqrt{-h} \, B_{MN} (X) \partial_a X^M \partial_b X^N \varepsilon^{ab} 
$$
We expect the latter term to shift $P_M = \frac{\partial L_G}{\partial \dot{X}^M}$, i.e. the momentum in absence of the coupling with the electric field. In particular, the gauge potential \eqref{eq:gaugepotential} and the embedding coordinates \eqref{eq:embedding} imply that 
\be \label{eq:LB}
L_B = \frac{n R_y}{\alpha'} \frac{Q}{r} \dot{X}^t
\ee
where the integral in the coordinate $\sigma$ has been easily performed since no dependence on $\sigma$ is left in the integrand. Being \eqref{eq:LB} only dependent on $\dot{X}^t$ (and not on the other $\dot{X}$s) all conjugate momenta are left invariant by the coupling with the gauge field, except
\be
P_t \rightarrow P_t + \frac{q Q}{r}
\ee
or equivalently in terms of the energy $E$ ($P_t = -E$)
\be \label{eq:Eshift}
E \rightarrow E - \frac{q Q}{r}
\ee
In these expressions we have implicitly used the definition  
\be \label{eq:qdefinition}
q = \frac{n R_y}{\alpha'}
\ee

The presence of a compact direction also leads to a shift in the mass of the probe (compared to point-like particles in 4 dimensions): 
\be \label{eq:massshift}
\mu^2 \rightarrow \mu^2 + f_b(r) q^2 + \frac{p^2}{f_b (r)} 
\ee
The warping factor \eqref{eq:warping} $f_b = 1 - \frac{r_b}{r}$ arises from the metric \eqref{eq:top}: indeed, the mass of the string depends on the actual length of the compactification radius which is modulated by the $g_{yy}$ component of the metric. In \eqref{eq:massshift}, besides \eqref{eq:qdefinition}, we have made use of the definition 
\be
p = \frac{\hbar m}{R_y}
\ee
where $m$ is the integer number representing Kaluza-Klein momentum units.

\subsection{Charged scalars in the Topological Star background}
At the linear level, scalar perturbations decouple from vector and tensor modes. Therefore, the wave equation for a massive scalar in the background of a TS is simply the Klein-Gordon equation\footnote{Keep in mind that, according to the measure units of Appendix \ref{appmeas}, the mass parameter in this equation is actually $\tilde{\mu}^2 = \frac{\mu^2}{\hbar^2}$. This also applies to the momentum $p$ appearing in \eqref{eq:waveeq}: it should be $\tilde{p} = p/\hbar$. In the following we will omit the tilde in both instances.}
\be
\Box \Phi = \mu^2 \Phi
\ee
The radial and angular dynamics can be separated by imposing the ansatz
\be \label{eq:waveeq}
\Phi=e^{i(-\omega t+p y+m_\phi\phi)}R(r)S(\q)
\ee
In this expression, the exponential contains the conserved conjugate momenta: for instance $\omega$ is linked to the energy introduced in \eqref{eq:Eshift}, through the definition $E = \hbar \omega$.\\The solutions $S(\q)$ of the angular equation are the spherical harmonics in $S^2$. The radial equation, instead, reads
\be
\label{eqradial}
(r-r_b)(r-r_s)R''(r)+(2r-r_b-r_s)R'(r)+\Bigg[r^2\left(\frac{r\omega^2}{r-r_s}-\frac{r p^2}{r-r_b}-\mu^2\right)-\ell(\ell+1)\Bigg]R(r)=0
\ee

The charged scalar modes analyzed in this paper can be seen as scalar excitations of the stringy probes of Subsection \ref{Stringy}. The coupling to the electric field and the string nature of the probe can be taken into account by replacing the frequency and the mass according to \eqref{eq:Eshift} and \eqref{eq:massshift}. Thus, in the radial equation we perform the following shifts
\be\label{shift}
\omega\rightarrow \omega-\frac{q Q}{r} \quad,\quad \mu^2\rightarrow \mu^2+q^2 \left(1-\frac{r_b}{r}\right) + \frac{p^2} {(1 - \frac{r_b}{r})}
\ee
After these substitutions, the radial equation \eqref{eqradial} can be recast in the canonical (Schr\"odinger) form 
\be
\label{eq:eqradialcan}
\psi''(r)+Q_W(r) \psi(r)=0
\ee
by introducing the function $\psi(r)$, related to the original one by
\be
R(r)=\frac{\psi(r)}{\sqrt{(r-r_s)(r-r_b)}}
\ee
In our instance, the characteristic $Q-$function $Q_W(r)$ appearing in \eqref{eq:eqradialcan} is
\begin{align}
\label{eq:Qgrav1}
Q_W(r)=&\frac{A_4 r^4+A_3 r^3+A_2 r^2+ A_1 r+A_0}{4(r-r_b)^2(r-r_s)^2}\\\nn
A_4=&-4(\omega^2-\m^2-p^2-q^2)\\\nn
A_3=&4(p^2 r_s + q^2 (2 r_b + r_s) + (r_b + r_s)\m^2 - 2 q Q\omega - r_b\omega^2)\\\nn
A_2=&4(\ell(\ell+1)+q^2(-Q^2+r_b^2+2r_br_s)+r_br_s\m^2-2q Qr_b\omega)\\\nn
A_1=&4(\ell(\ell+1)(r_b+r_s)+q^2r_b(r_br_s-Q^2))\\\nn
A_0=&r_b^2-2(1+2\ell(\ell+1))r_br_s+r_s^2
\end{align}
From the expression \eqref{eq:Qgrav1} of the $Q$-function, we infer that the radial dynamics of charged scalar perturbations in TS background is described by a Confluent Heun Equation (CHE); this is a Fuchsian differential equation with two regular singularities at $r=r_b$ and $r=r_s$ and an irregular one at $r= \infty$.



\section{Quantum Seiberg-Witten curves for \texorpdfstring{$\mathcal{N}=2$}{N=2} SYM with flavours} \label{quantumSW}
In this section, following the notation of \cite{Bianchi:2021mft}, we provide a short overview of how to construct quantum Seiberg-Witten (qSW) curves for $\mathcal{N}=2$ theories with an $SU(2)$ gauge group. Subsequently, we discuss how to compute the quantum periods $(a, a_D)$, which are instrumental in deriving exact quantization conditions for QNM frequencies and characterize the response of the geometry to tidal deformations.
\subsection{Quantum Seiberg-Witten curves}
The classical SW curves for $SU(2)$ gauge theory with $N_f=3$ fundamental hypermultiplets with masses $m_i$ in flat space are:
\be
\Lambda^2 y^2 P_L(x)+y P_0(x)+P_R(x)=0
\ee
with
\be\label{classSW}
P_L(x) {=} x{-}m_3, \qquad P_0(x) {=} x^2{-}u{+}\Lambda \left(x{+}\frac{1}{2}{-}m_1{-}m_2{-}m_3\right), \qquad P_R(x) {=} (x{-}m_1) (x{-}m_2)
\ee
In these definitions $\Lambda=e^{2\pi i \tau}$, $\tau$ is the gauge coupling and $u$ is the Coulomb branch (CB) modulus. In presence of a non trivial $\Omega-$background, $\epsilon_1=1$, $\epsilon_2=0$, the dynamics of the gauge theory is described by the quantum curve which can be obtained by the classical one by promoting the variables $x$ and $y$ to operators satisfying the commutation relation:
\be
[\hat{x},\ln{\hat{y}}]=1
\ee
The qSW curve follows from the classical one \eqref{classSW} 
\be
\label{eqnSWorig}
\left[\Lambda \hat{y}^{1/2}  P_L(\hat{x})\hat{y}^{1/2} +  P_0(\hat{x}) + \hat{y}^{-1/2}P_R (\hat{x})\hat{y}^{-1/2}\right] f(z)=0
\ee
This gauge theory can be understood in terms of D4-branes forming pairs stretched between two {NS5}-branes or between an NS5 and infinity, positioned both to the left and right. The locations of the zeros of $P_{L,R}(x)$, $P_0(x)$ determine the positions of the D4-branes in these three regions, serving as parameters for the four masses and defining the CB.

Equation \eqref{eqnSWorig} can be put in normal form by redefining the wave function as 
\be
f(z) = e^{-\frac{q z}{2}} z^{\frac{m_1+m_2}{2}} (1+z)^{-\frac{1+m_1+m_2}{2}} \Psi(z)
\ee
in such a way that \eqref{eqnSWorig} becomes
\be
\label{eqnSWcan}
\Psi''(z) + Q_{12}(z) \Psi(z) = 0
\ee
with
\be
\label{QSW12}
Q_{12}(z) {=}{-}\frac{\L^2}{4}{+}\frac{1{-}(m_1{-}m_2)^2}{4z^2}{+}\frac{1{-}(m_1{+}m_2)^2}{4(1{+}z)^2}{-}\frac{m_3\Lambda}{z}{-}\frac{1{-}2m_1^2{-}2m_2^2{+}4u{+}2\Lambda(m_1{+}m_2{-}1)}{4z(1{+}z)}
\ee

\subsection{Gauge/gravity dictionaries}
The identification between $Q_W$ \eqref{eq:Qgrav1} and $Q_{12}$ \eqref{QSW12} allows us to construct the dictionary between the parameters of the gravity solution and of the gauge theory side. 
For the TS we choose:
\be
z = \frac{r-r_b}{r_b-r_s}
\ee
in such a way that $r= \infty \leftrightarrow z = \infty$, $r=r_b \leftrightarrow z=0$ and $r=r_s \leftrightarrow z=-1$. This is suitable for the present case because the values $z<0$ are not physical, exactly like for the case with $r<r_b$. 
By doing this change of variables, and setting $\mu=p=0$ for simplicity, we get the following dictionary
\begin{align}
\label{dictionary ts}
&\hspace{1.5cm} m_1 =m_2= -\frac{(q Q-r_s\omega)}{\sqrt{\frac{r_b}{r_s}-1}},\quad  m_3=\frac{-q^2 r_s - 2 q Q \omega + (r_b+ 2 r_s) \omega^2}{2i \sqrt{\omega^2-q^2}} \\
& u = \left(\ell{+}\frac{1}{2} \right)^2{-}q^2[Q^2{+}(r_b{-}r_s)r_s]{+}4 q Q r_s \omega {-} 3 r_s^2 \omega^2 {-}i\sqrt{\omega^2{-}q^2}(r_b{-}r_s)\left(1{+}2\frac{q Q{-}r_s\omega}{\sqrt{\frac{r_b}{r_s}{-}1}}\right) \notag\\ 
&\hspace{5cm}\Lambda= -2 \text{i} (r_b-r_s) \sqrt{\omega^2-q^2} \notag 
\end{align}

We choose to impose $p=\mu=0$ also for the BS. In this instance, 
\be
z=\frac{r-r_s}{r_s-r_b}
\ee
 so that the dictionary is the following:
  \be\label{dictBS}
m_1=-m_2=\frac{i(r_s\omega-q Q)}{\sqrt{1-\frac{r_b}{r_s}}},\quad m_3=\frac{-q^2 r_s -2q Q\omega+(r_b+2r_s)\omega^2}{2i\sqrt{\omega^2-q^2}}
  \ee
 $$ 
  u=\left(\ell+\frac{1}{2}\right)^2-q^2Q^2+2qQ(r_b+r_s)\omega-\omega^2(r_b^2+r_b r_s+r_s^2)+i (r_b-r_s)\sqrt{\omega^2-q^2}
$$
 $$\Lambda=-2i(r_s-r_b)\sqrt{\omega^2-q^2}$$

\subsection{Quantum Seiberg-Witten periods}
The dynamics of the gauge theory is codified in terms of a single holomorphic function which is the qSW period $a$. In order to compute it, we have to Fourier transform \eqref{eqnSWorig} bringing it in the form of a difference equation \cite{Poghossian:2010pn,Fucito:2011pn}, whose integrability condition can be written as a continued fraction:
\be\label{diffeqSW}
P_0(a)=\frac{M(a+1)}{P_0(a+1)-\frac{M(a+2)}{P_0(a+2)-...}}+\frac{M(a)}{P_0(a-1)-\frac{M(a-1)}{P_0(a-1)-...}}
\ee
with
\be
M(x)=\Lambda P_L\left(x-\frac{1}{2}\right)P_R\left(x-\frac{1}{2}\right)
\ee
Equation \eqref{diffeqSW} with \eqref{classSW} can be solved for $u(a,\Lambda)$ order by order in $\Lambda$, getting
\begin{align}
\label{uvsa}
u(a,\Lambda) = & a^2 + \Lambda \left[\frac{1}{2}(1-m_1-m_2-m_3) - \frac{2 m_1 m_2 m_3}{4 a^2 - 1}\right] + \frac{\Lambda^2}{128 (a^2-1)} \left[4 a^2 -5 + \notag \right.\\
&\left.+ 4(m^2_1+m^2_2+m^2_3) - \frac{48 (m_1^2 m_2^2 + m_2^2 m_3^2 + m_1^2 m_3^2)}{4a^2 -1} + \frac{64 m^2_1 m^2_2 m^2_3(20 a^2+7)}{(4 a^2 -1)^3} \right]+\notag\\&+\dots
\end{align}
We can invert this series in order to obtain 
\begin{align}
\label{avsu}
a(u,\Lambda) =& \sqrt{u} + \frac{\Lambda}{4 \sqrt{u}} \left(\frac{4 m_1 m_2 m_3}{4u-1} + m_1 + m_2 + m_3 - 1\right) + \\\nn
&-\frac{\Lambda ^2}{256 \sqrt{u}} \left(\frac{1024 m_1^2 m_2^2 m_3^2}{(4
   u-1)^3}-\frac{256 m_1 m_2 \left(m_1 \left(m_2 m_3-2\right)-2
   \left(m_2+m_3-1\right)\right) m_3}{(1-4 u)^2}+\right.\\\nn
   &\left.+\frac{8
   \left(m_2+m_3+m_1 \left(1-4 m_2
   m_3\right)-1\right){}^2}{u}+\frac{\left(4 m_1^2-1\right)
   \left(4 m_2^2-1\right) \left(4 m_3^2-1\right)}{u-1}+\right.\\\nn
   &\left.+\frac{64
   \left(\left(\left(1{-}12 m_3^2\right) m_2^2{+}4 m_3
   m_2{+}m_3^2\right) m_1^2{+}4 m_2 m_3 \left(m_2{+}m_3{-}1\right)
   m_1{+}m_2^2 m_3^2\right)}{4 u{-}1}{+}4\right)+\\&{+}\dots
\end{align}
The Nekrasov-Shatashvili (NS) prepotential $\mathcal{F}_{NS}(a,\Lambda)$ is obtained by integrating the quantum version of the Matone relation \cite{Matone:1995rx,Flume:2004rp}
\be\label{matonerel}
u=-\Lambda \frac{\partial \mathcal{F}_{NS}(a,\Lambda)}{\partial \Lambda}
\ee
Furthermore, we have to add a $\Lambda$-indipendent constant obtained from the one-loop prepotential, so that
\be
\mathcal{F}_{NS}(a,\Lambda)=\mathcal{F}_{tree}(a,\Lambda)+\mathcal{F}_{1-loop}(a)+\mathcal{F}_{inst}(a,\Lambda)
\ee
where
\begin{align}
\label{Prepotential}
& \mathcal{F}_{\text{tree}}(a,\Lambda) = -a^2 \log \Lambda \notag \\
& \mathcal{F}_{\text{inst}}(a,\Lambda) = \Lambda \frac{1-m_2-m_3+4 a^2 (m_1 + m_2 + m_3 -1) + m_1 (4 m_2 m_3 - 1)}{2(4a^2-1)} + \dots\\\nn
&\frac{\partial \mathcal{F}_{\text{one-loop}}(a)}{\partial a} = \log \left[\frac{\Gamma^2(1+2a)}{\Gamma^2(1-2a)} \prod_{i=1}^3 \frac{\Gamma\left(\frac{1}{2}+m_i-a\right)}{\Gamma\left(\frac{1}{2}+m_i+a\right)}\right]
\end{align}
Finally, the $a_D-$period is given by
\be\label{addef}
a_D=-\frac{1}{2\pi i} \frac{\partial \mathcal{F}_{NS}}{\partial a}
\ee
QNM frequencies are obtained by imposing the exact quantization condition \cite{Bonelli:2022ten,Consoli:2022eey}:
\be\label{condquantqnm}
a_D=n\quad,\quad n=0,1,2,...
\ee
once expressed in terms of the gravity variables via the dictionary. The latter is solved in $\omega$ in the complex plane.

\section{Leaver method} \label{Leaver}

In this section, we describe a semi-analytical method used to solve the radial differential equation involving a continued fraction, introduced by Leaver \cite{Leaver:1985ax, Leaver:1990zz}. This method has been applied in \cite{Bianchi:2023sfs, Heidmann:2023ojf} to the study of QNMs for neutral scalar waves in TS geometry. 

For generality, let us implement this technique on the (1,2)-SW curve \eqref{eqnSWcan}-\eqref{QSW12}, once we redefine:
\be
\Psi(z)=\sqrt{z(1+z)}R(z)
\ee
so that
\begin{equation}\label{12qswN}
z(1+z)R''(z)+(1+2z)R'(z)+\Big[-\frac{\Lambda^2z^2}{4}+\frac{\Lambda(\Lambda+4m_3)z}{4}-\frac{(m_1-m_2)^2}{4z}+\frac{(m_1+m_2)^2}{4(1+z)}+
\end{equation}
$$+\frac{1}{4}(1-4u+2\Lambda(1-m_1-m_2-2m_3))\Big]R(z)=0$$
The coordinate $z$ is the coordinate that originates at the cap in the case where the geometry is smooth $(r_b>r_s)$, while it originates at the horizon in the case of a BS:
\be
z = \begin{cases}
\frac{r - r_b}{r_b - r_s} & \quad \mathrm{if} \ r_b>r_s \\
\frac{r - r_s}{r_s - r_b} & \quad \mathrm{if} \ r_s > r_b
\end{cases}
\ee
Once the appropriate gauge/gravity dictionary is chosen, the Frobenius exponents for the regular singular points are chosen in such a way that at $z=0$ they encode the information that the solution is regular in the case of smooth geometry or that the wave is ingoing in the case of a BS, while at infinity, where the singularity is irregular, the wave is outgoing in both cases. The other Frobenius exponent at $z=-1$ is chosen such that the recurrence relation is three-termed. Therefore, the correct ansatz is:
\begin{equation}\label{leava}
    R_L(z)=e^{-\frac{\Lambda z}{2}}z^{-\frac{m_1-m_2}{2}}(1+z)^{\frac{m_1-m_2-2(1+m_3)}{2}}\sum_{n=0}^\infty c_n\left(\frac{z}{1+z}\right)^n
\end{equation}
If we plug \eqref{leava} in \eqref{12qswN}, we obtain a three terms recursion relation of the form:
\be
\alpha_{n}c_{n+1}+\beta_n c_n+\gamma_n c_{n-1}=0\quad,\quad n\ge 0
\ee where the coefficients are
\begin{align}
\label{coeff Leaver}
\alpha_n=&(1+n)(1+n-m_1+m_2)\\\nn
\beta_n=&{-}\left(m_3{+}n\right) (\Lambda {+}2 n){-}m_2 \left(\Lambda {+}m_3{+}2
   n{+}1\right){+}m_1 \left(m_2{+}m_3{+}2 n{+}1\right){-}m_3{-}2
   n{-}u{-}\frac{3}{4}\\\nn
\gamma_n=&(n+m_2+m_3)(n-m_1+m_3)
\end{align}
and where we chose $c_{-1}=0$. The QNMs associated to the overtone $n$ are obtained by truncating the recursion and solving numerically the continuous fraction:
\be
\label{fraction Leaver}
\beta_n =\frac{\alpha_n\gamma_n}{\beta_{n-1}-\frac{\alpha_{n-2}\gamma_{n-1}}{\beta_{n-2}-\dots}}+\frac{\alpha_n\gamma_{n+1}}{\beta_{n+1}-\frac{\alpha_{n+1}\gamma_{n+2}}{\beta_{n+2}-\dots}}
\ee
viewed as an equation for the QNMs $\omega$.

The two methods exposed in Section \ref{quantumSW} and in the current one can be applied in both cases of TSs ($r_b>r_s$) and BSs ($r_s>r_b$). 

\section{Charged QNMs of Topological Star} \label{TopstarQNM}
In this Section we analyze the case of TSs. We compute the QNMs with qSW technique and Leaver method; furthermore, we obtain the same quantities also with a direct integration and then we compare the results obtained with these different methods.
\subsection{Computation of QNMs}
\begin{figure}[H]
    \centering
    \includegraphics[width=0.5\textwidth]{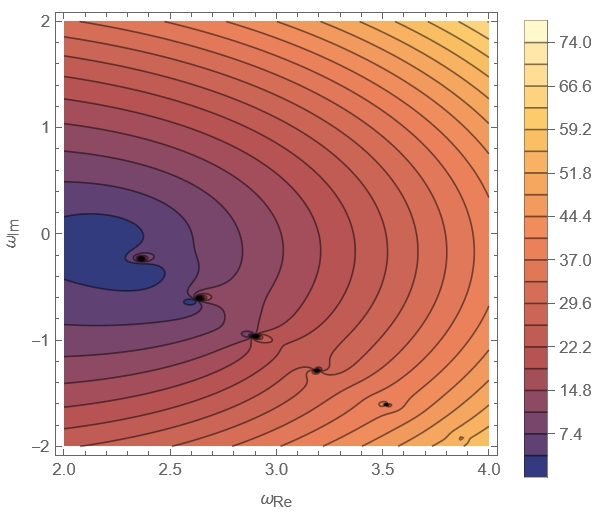} \\
    \includegraphics[width=0.3\textwidth]{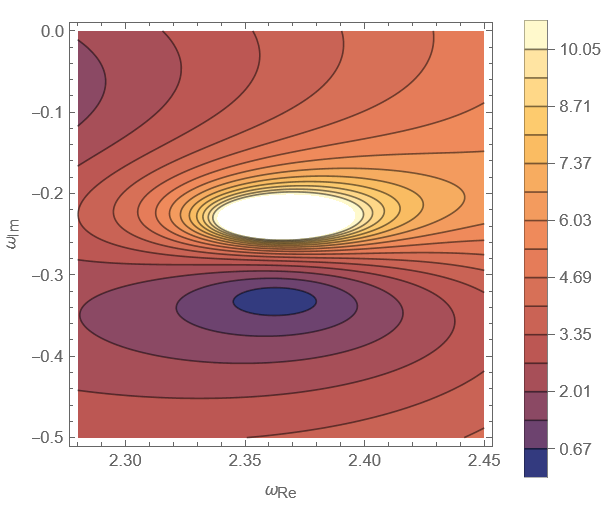}
    \includegraphics[width=0.3\textwidth]{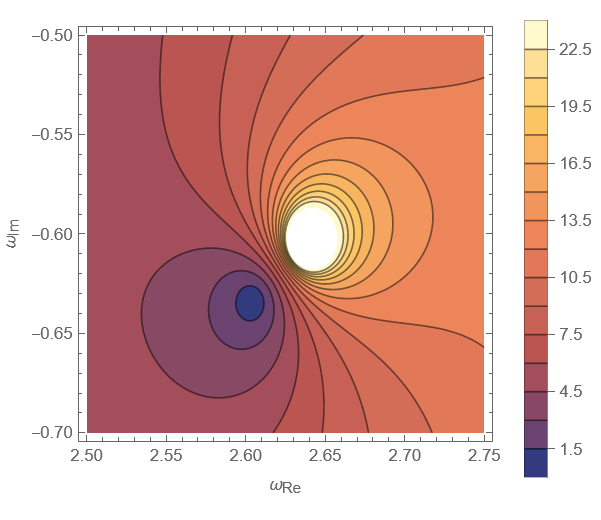}
    \includegraphics[width=0.3\textwidth]{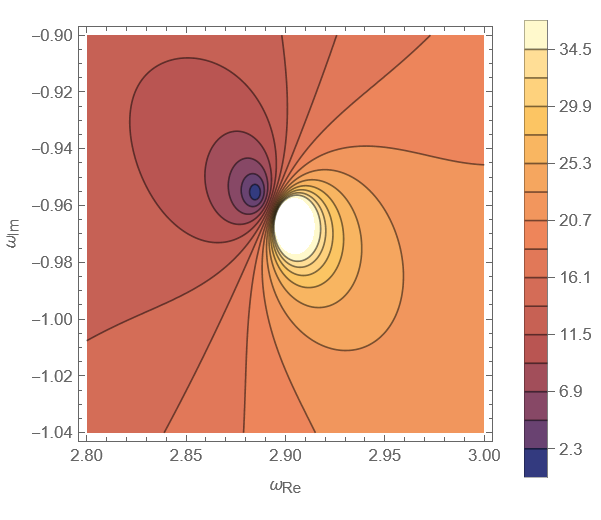}
    \label{fig:immagine1}
    \caption{Plots of the modulus of the Leaver difference equation \eqref{fraction Leaver} in the complex $\omega$-plane for $r_b=1$, $r_s=0.8$, $\ell=2$, $Q=2$, $q=0.5$, $p=0$, $\m=0$. The three plots at the bottom are zoomed-in snapshots of the regions around the first three zeros of the Leaver function.}\label{plotL}
\end{figure}

The dictionary \eqref{dictionary ts} is the fundamental ingredient required for implementing the Leaver method \eqref{coeff Leaver}, \eqref{fraction Leaver} and the SW quantization condition \eqref{Prepotential}, \eqref{addef}, \eqref{condquantqnm}. As demonstrated in Figure \ref{plotL} and Tables \ref{Tabb1}, \ref{Tabb2} and \ref{Tabb3}, only stable modes are present. In the last column, by $SW_s$ we mean that the number of instantonic corrections used in the SW procedure is equal to $s$.

The resulting QNM frequencies have also been validated against those obtained using a direct integration method, which involves matching numerical solutions in different integration regions. Starting from the cap, in accordance with the solution $(\ref{RCAPts})$, we impose a regular boundary condition and integrate the equation up to a specified extraction point. From this point, following the solution $(\ref{RINF})$, we impose outgoing waves and continue integration until reaching infinity. The two solutions can be matched by requiring the vanishing of the Wronskian, which then becomes an eigenvalue equation for QNM frequencies. In Table \ref{Tabb2}, the column associated to the numerical integration is empty because this algorithm fails to be predictive when the QNMs are highly damped, that is when their imaginary parts are close to 1. Indeed in this scenario we need a huge number of subleading corrections at the cap and at infinity to estimate QNMs with high imaginary parts.

As it can be seen, the results of the QNMs corresponding to the different methods are in agreement. We expect that a larger number of instantonic corrections in the qSW method will produce estimates closer to the ones provided by numerical integration and Leaver method.

\begin{table}[H]
\centering
\begin{tabular}{|c|c|c|c|}
\hline
$\ell$ & Numerical integration & Leaver                & $SW_4$ \\ \hline
$0$    & $1.34819 - 0.218003 \, i$    & $1.34819 - 0.218003 \, i$  & $1.33277 - 0.261032 \, i$   \\ \hline
$1$    & $1.73838 - 0.126107 \, i$    & $1.73838 - 0.126107 \, i$  & $1.72053 - 0.117156 \, i$   \\ \hline
$2$    & $2.18231 - 0.0797013 \,  i$    & $2.18231 - 0.0797013 \, i$ & $2.17005 - 0.0762405 \, i$   \\ \hline
$3$    & $2.63417 - 0.0517219 \,  i$    & $2.63417 - 0.0517219 \, i$ & $2.6258 - 0.060272 \, i$   \\ \hline
\end{tabular}
\caption{TS QNMs for different values of $\ell$ and fixed overtone $n=0$ and $r_b=1$, $r_s=0.8$, $Q=2$, $q=0.5$, $p=\mu=0$.}\label{Tabb1}
\end{table}
\begin{table}[H]
\centering
\begin{tabular}{|c|c|c|c|}
\hline
$n$ & Numerical integration & Leaver                & $SW_2$ \\ \hline
$1$    & $-$    & $2.36322 - 0.330499 \, i$  & $2.24082 - 0.293755 \, i$   \\ \hline
$2$    & $-$    & $2.60406 - 0.633878 \, i$  & $2.83614 - 0.700827 \, i$   \\ \hline
$3$    & $-$    & $2.88498 - 0.955453 \, i$  & $2.89907 - 0.926271 \, i$   \\ \hline
\end{tabular}
\caption{TS QNMs for fixed $\ell=2$, $r_b=1$, $r_s=0.8$, $Q=2$, $q=0.5$, $p=\mu=0 $ and for various overtone numbers.}\label{Tabb2}
\end{table}
\begin{table}[H]
\centering
\begin{tabular}{|c|c|c|c|}
\hline
$n$ & Numerical integration & Leaver                & $SW_4$ \\ \hline
$0$    & $5.81013 - 0.000491187 \,i$    & $5.81013 - 0.000491187\, i$  & $5.81895 - 0.0139528 \,i$   \\ \hline
$1$    & $5.97759 - 0.0411436 \,i$    & $5.97759 - 0.0411436 \,i$  & $5.96178 - 0.0271379\, i$   \\ \hline
$2$    & $6.13468 - 0.17184\, i$    & $6.13468 - 0.17184\, i$  & $6.12144 - 0.14779\, i$   \\ \hline
$3$    & $-$    & $6.31023 - 0.343684 \, i$  & $6.30019 - 0.385102\, i$   \\ \hline
\end{tabular}
\caption{TS QNMs for fixed $\ell=10$, $r_b=1$, $r_s=0.8$, $Q=2$, $q=0.5$, $p=\mu=0 $ and for various overtone numbers.}\label{Tabb3}
\end{table}

\section{Charged QNMs and Superradiance of Black String} \label{Charge superradiance}
In this Section we pass to examine the BSs. We explain why BSs can produce charge superradiance and we also show that in TSs case, instead, this phenomenon cannot take place. At this point, we can introduce the \textit{near extremal, near superradiant limit} together with the \textit{probe limit}, arriving at computing QNMs with the usual two methods of qSW and Leaver and now also with the matching asymptotic expansion. Prompt Ring Down modes are expected as well and indeed they are computed, in addition to the two persistent methods, with a WKB approximation. At the end of this Section, the amplification factor of superradiance for BSs is computed, keeping in count also the first instantonic corrections.    
 
\subsection{Charge superradiance}
TS and BS are static and charged geometries and hence they can only give rise to charge superradiance. Focusing on BS, we have to determine first of all the behaviour of the radial wave function $\Psi(z)$ at the boundaries of the spacetime under consideration, that is $z=0$ and $z=\infty$. By solving \eqref{eqnSWcan} with $Q_{12}$ given by \eqref{QSW12} adapted at the two boundaries in exam, we get
\be
\Psi(z) \simeq 
\begin{cases}
 \mathcal{O} z^{\frac{1}{2} - i \, \, \frac{q Q - \omega r_s}{\sqrt{1-\frac{r_b}{r_s}}}} +  \mathcal{T} z^{\frac{1}{2} + i \, \, \frac{q Q - \omega r_s}{\sqrt{1-\frac{r_b}{r_s}}}}, \qquad\qquad\qquad z=0 \\
 \mathcal{R} e^{ i (r_s-r_b) \, z \, \,\sqrt{\omega^2-q^2}} +  \mathcal{I} e^{ -i (r_s-r_b) \, z \, \,\sqrt{\omega^2-q^2}}, \quad z \to +\infty\\
\end{cases}
\ee
where, following the convention of \cite{Brito:2015oca}, $\mathcal{I}$ and $\mathcal{R}$ are, respectively, the amplitudes of  incident and reflected waves (defined at spatial infinity $z = \infty$), while $\mathcal{T}$ and $\mathcal{O}$ are the amplitudes of transmitted and outgoing waves across the boundary $z=0$ (the horizon, in this geometry).\\
The probability currents associated to the two asymptotic solutions are:
\begin{align}
& J_0 = \frac{2 i (-q Q + r_s \omega)}{\sqrt{1-\frac{r_b}{r_s}}} (|\mathcal{O}|^2-|\mathcal{T}|^2) \\
& J_\infty = 2 \, i \, (r_b-r_s) \sqrt{\omega^2-q^2} (|\mathcal{I}|^2-|\mathcal{R}|^2)
\end{align}
and they must be equal due to the conservation and this implies:
\be
\label{incident wave amplitude}
|\mathcal{I}|^2 = |\mathcal{R}|^2 + \frac{(q Q-r_s \omega) \sqrt{r_s}}{(r_s-r_b)^{\frac{3}{2}} \sqrt{\omega^2-q^2}} (|\mathcal{O}|^2 - |\mathcal{T}|^2) 
\ee
Notice that if $\omega < \omega_{SR} = \frac{q Q}{r_s}$ then the wave is superradiantly amplified ($|\mathcal{I}|^2 < |\mathcal{R}|^2$) \footnote{For a BH the boundary conditions at the horizon involve only ingoing waves, so that $\mathcal{O}=0$.}. For this reason $\omega_{SR}$ will be denoted as \textit{superradiance threshold}.\\
This result is in completely agreement with the one we could obtain by using the classical laws of BH thermodynamics, as firstly understood by Bekenstein \cite{PhysRevD.7.949}. Indeed, the reduction from 5 to 4 dimensions for BS gives rise to a RN BH with radius of the horizon equal to $r_s$. The first law of thermodynamics for this kind of BH reads
\be
\label{1st law BH}
\delta M = \frac{T_H}{4} \delta A_H + \Phi_H \delta \mathcal{Q} 
\ee
where $T_H$, $A_H$, $\Phi_H$ are the temperature, the area and the electric potential at the horizon, hence $\Phi_H = \mathcal{Q}/r_s$. If we send a monochromatic wave with frequency $\omega$ and charge $q$, the variation of the mass of the background in terms of its electric charge is
\be
\delta M = \delta \mathcal{Q} \frac{\omega}{q}
\ee
and so, plugging in \eqref{1st law BH}, we get
\be
\delta \mathcal{Q} = \frac{T_H q}{4} \frac{\delta A_H}{\omega - q \Phi_H}
\ee
In order to have charge superradiance ($\delta \mathcal{Q} < 0$), since $\delta A_H \geq 0$ for the second law of BH thermodynamics, we must have $\omega - q \Phi_H < 0$. This is exactly what we found out in \eqref{incident wave amplitude}.

It is worth noticing that in the case of TS $(r_b>r_s)$, it is not possible to have superradiant amplification. This circumstance arises from the fact that smooth geometries resolve the information paradox and therefore the amplitude of the incident wave on our compact object will be equal to the reflected one. Once the regularity is imposed at the cap of the geometry $r=r_b$, the amplification curve obtained by numerically integrating the radial wave equation is zero. This will be shown in Figure \ref{plotAmpli} together with the amplification factor of BS.

\subsection{Near extremal and near superradiant modes}
In this subsection we will analytically point out some peculiar modes, referred as \textit{near extremal, near superradiant modes}, that will be computed by exploiting some matching asymptotic expansion technique.\\
Firstly notice that in the near extremal limit ($r_s \to r_b$) the theory decouples ($\Lambda \to 0$) and, in order to avoid divergences arising in the dictionary \eqref{dictBS}, we must fix $\omega = \omega_{SR}$. In this way the dictionary becomes:
\be
m_1=m_2=\Lambda=0,\quad u=\left(\ell+\frac{1}{2}\right)^2,\quad m_3=-\frac{i q}{2}\sqrt{Q^2-r_s^2}
\ee
In order to reproduce results known in literature \cite{Consoli:2022eey,Starobinskil:1974nkd}, we consider the approximation $m_3 \simeq 0$, which is consistent with $q \simeq 0$ \footnote{This parameter can be small since it is proportional to $R_y$.}, called \textit{probe limit}.

Let us now apply the technique previously cited.
The differential equation \eqref{eqnSWcan} with $Q_{12}$ in \eqref{QSW12} in the decoupling limit $\Lambda=0$ can be solved in terms of hypergeometric functions. The solution that guarantees ingoing waves at the horizon $z=0$ is
\be
\Psi(z)=z^{\frac{1}{2}(1-m_1+m_2)}(1+z)^{\frac{1}{2}(1+m_1+m_2)}{}_2F_1\Big[\frac{1}{2}+m_2-\sqrt{u},\frac{1}{2}+m_2+\sqrt{u},1-m_1+m_2,-z\Big]
\ee
The behaviour of the previous solution at infinity is
\be\label{sol1}
\Psi(z)\sim z^{\frac{1}{2}-\sqrt{u}}\left(1+{z^{2\sqrt{u}}\Gamma(\frac{1}{2}-m_1-\sqrt{u})\Gamma(\frac{1}{2}+m_2-\sqrt{u})\Gamma(2\sqrt{u})\over \Gamma(\frac{1}{2}-m_1+\sqrt{u})\Gamma(\frac{1}{2}+m_2+\sqrt{u})\Gamma(-2\sqrt{u})}\right)
\ee
At $z=\infty$ the differential equation becomes
\be
\Psi''(z)+\left(-{m_3\Lambda\over z}-{\Lambda^2\over4}+{1-4u\over 4z^2}\right)\Psi(z)=0
\ee
whose general solution is
\be\label{solinfpm}
\Psi(z)=\sum_{\alpha=\pm}A_\alpha e^{{\alpha\Lambda z\over2}}(z\Lambda)^{{1\over2}{+}\sqrt{u}}U\Big[{1\over2}{-}\alpha m_3{+}\sqrt{u},1{+}2\sqrt{u},{-}\alpha\Lambda z\Big]
\ee
In order to have only outgoing waves at infinity, we are forced to set $A_+=0$ (and $A_-=1$ without loss of generality). After that we can expand the previous \eqref{solinfpm} solution at the horizon
\be\label{sol2}
\Psi(z)\sim z^{{1\over2}-\sqrt{u}}\left(1+{z^{2\sqrt{u}}\Lambda^{2\sqrt{u}}\Gamma({1\over2}+m_3+\sqrt{u})\Gamma(-2\sqrt{u})\over \Gamma\left(\frac{1}{2}+m_3-\sqrt{u}\right)\Gamma(2\sqrt{u})}\right)
\ee
The two solutions \eqref{sol1} and \eqref{sol2} match if:
\be\label{1loop}
\Lambda^{2\sqrt{u}}-{\Gamma(1+2\sqrt{u})^2\Gamma({1\over2}-m_1-\sqrt{u})\Gamma\left(\frac{1}{2}+m_2-\sqrt{u}\right)\Gamma\left(\frac{1}{2}+m_3-\sqrt{u}\right)\over \Gamma(1-2\sqrt{u})^2\Gamma\left(\frac{1}{2}-m_1+\sqrt{u}\right)\Gamma\left({1\over2}+m_2+\sqrt{u}\right)\Gamma\left(\frac{1}{2}+m_3+\sqrt{u}\right) }=0
\ee
Since the first term in \eqref{1loop} is going to zero, in order to satisfy the equation we require that one of the $\Gamma-$functions in the denominator of the second term must diverge. The correct choice is
\be
{1\over2}+m_2+\sqrt{u}=-n\quad, \quad n=0,1,2,...
\ee
Using the dictionary \eqref{dictBS} and replacing $\omega=\omega_{SR}-\nu \sqrt{r_s-r_b}$, we can find the correction to the superradiant frequency by expanding for small $r_s-r_b$:
\be
\label{omega match}
\omega_{match}={q Q\over r_s}-i{\ell+1+n\over r_s^{3/2}}\sqrt{r_s-r_b}
\ee
The observant reader will have noticed that the matching condition \eqref{1loop} coincides with the quantization of the cycle $a_D$ as written in \eqref{condquantqnm}, where only the tree-level and the 1-loop contributions are present. So, starting from \eqref{1loop}, in order to consider the istantonic corrections, let us replace $\sqrt{u}=a$ with the $a$-cycle defined as \eqref{avsu} and we also add the instantonic contribution of the prepotential \eqref{Prepotential}, getting at the end
\be
\label{SW freq}
\Lambda^{2a}-{\Gamma(1+2a)^2\Gamma({1\over2}-m_1-a)\Gamma(\frac{1}{2}+m_2-a)\Gamma(\frac{1}{2}+m_3-a)\over \Gamma(1-2a)^2\Gamma(\frac{1}{2}-m_1+a)\Gamma({1\over2}+m_2+a)\Gamma(\frac{1}{2}+m_3+a) }e^{\partial_a F_{inst}(\Lambda)}=0
\ee 
The results obtained with the matching asymptotic expansion \eqref{omega match}, with SW quantization \eqref{SW freq} and with Leaver continued fraction \eqref{coeff Leaver},\eqref{fraction Leaver}, using the dictionary \eqref{dictBS}, are displayed in Tables \ref{Tab 2 BS} and \ref{Tab 3 BS}.

\begin{table}[H]
\centering
\begin{tabular}{|c|c|c|c|}
\hline
$r_b$   & $\omega_{match}$               & Leaver                 & $SW_4$                 \\ \hline
$0.9$   & $1 - 0.948683 \, i$    & $1.51407 - 1.10179 \, i$  & $1.51404 - 1.10182 \, i$  \\ \hline
$0.95$  & $1 - 0.67082 \, i$     & $1.06253 - 0.7356 \, i$   & $1.06253 - 0.7356 \, i$   \\ \hline
$0.99$  & $1 - 0.3 \, i$        & $1.01209 - 0.305512 \, i$ & $1.01209 - 0.305512 \, i$ \\ \hline
$0.999$ & $1 - 0.0948683 \, i$ & $1.0012 - 0.0950405 \, i$ & $1.0012 - 0.0950407 \, i$ \\ \hline
\end{tabular}
\caption{Near extremal near superradiant modes corresponding to the first overtone number $n=0$ for $r_s=1$, $\ell=2$, $q=0.5$, $Q=2$, $\mu=p=0$.}
\label{Tab 2 BS}
\end{table}

\begin{table}[H]
\centering
\begin{tabular}{|c|c|c|c|}
\hline
$r_b$   & $\omega_{match}$               & Leaver                 & $SW_4$                 \\ \hline
$0.9$   & $1 - 1.26491 \, i$    & $1.4123 - 1.39 \, i$  & $1.41232 - 1.38995 \, i$  \\ \hline
$0.95$  & $1 - 0.894427 \, i$     & $1.07965 - 1.00857 \, i$   & $1.07965 - 1.00857 \, i$   \\ \hline
$0.99$  & $1 - 0.4 \, i$        & $1.01595 - 0.409877 \, i$ & $1.01596 - 0.409876 \, i$ \\ \hline
$0.999$ & $1 - 0.126491 \, i$ & $1.0016 - 0.126801 \, i$ & $1.0016 - 0.126801 \, i$ \\ \hline
\end{tabular}
\caption{Near extremal near superradiant modes corresponding to the second overtone number $n=1$ for $r_s=1$, $\ell=2$, $q=0.5$, $Q=2$, $\mu=p=0$.}
\label{Tab 3 BS}
\end{table}

\subsection{Prompt Ring Down modes}

As extensively shown in \cite{Bianchi:2022qph,Bianchi:2023sfs,DiRusso:2024hmd}, in general, within smooth and horizonless solutions, one can distinguish between unstable and metastable modes, corresponding to frequencies near the maximum or minimum of the effective potential, respectively. Unstable frequencies are also referred to as prompt ring down modes, bearing in mind the analogy with BH mergers. Given the shape of the effective potential of the BS, which exhibits only one unstable light ring, we expect such modes to be present in the spectrum of QNMs and we aim to provide a WKB estimate of their frequency values. Exploiting the results shown in \cite{Bianchi:2021mft}, prompt ring down frequencies must satisfy the Bohr-Sommerfeld quantization condition
\be
\int_{z_-}^{z_+}\sqrt{Q(z)}dz=\pi\left(n+\frac{1}{2}\right)
\ee
with $n$ the overtone number and $z_{\pm}$ the two turning points. In the semiclassical limit the turning points collide and the integral can be approximated as
\be
\int_{z_-}^{z_+}\sqrt{Q(z)}dz\simeq \int_{z_-}^{z_+}\sqrt{Q(z_c)+\frac{Q''(z_c)}{2}(z-z_c)^2}dz\simeq\frac{i\pi Q(z_c)}{\sqrt{2Q''(z_c)}}
\ee
where $z_c\in[z_-,z_+]$ the critical point $Q'(z_c)=0$. So we are left with
\be
\frac{Q(z_c)}{\sqrt{2Q''(z_c)}}=-i\left(n+\frac{1}{2}\right)
\ee
The equation can be solved perturbatively by substituting the information that $\omega=\omega_{\text{R}}+i \omega_{\text{I}}$ and expanding for small $\omega_{\text{I}}$. It is easy to see that $\omega_{\text{R}}$ is simply the critical frequency at the unstable photon sphere. The results are shown in Tables \ref{tabb6} and \ref{tabb7} and are compared with the results provided by the Leaver method and SW quantization.
\begin{table}[H]
\centering
\begin{tabular}{|c|c|c|c|}
\hline
$\ell$ & WKB                   & Leaver               & $SW_4$                    \\ \hline
$0$    & $0.884096 - 0.402014 \, i$ & $1.04019 - 0.582305 \, i$ & $1.0416 - 0.571598 \, i$ \\ \hline
$1$    & $1.31439 - 0.144228 \, i$  & $1.36621 - 0.13088 \, i$  & $1.37756 - 0.138392 \, i$  \\ \hline
$2$    & $1.68218 - 0.13711 \, i$   & $1.71123 - 0.133916 \, i$ & $1.71388 - 0.108494 \, i$  \\ \hline
$3$    & $2.05574 - 0.135935 \, i$  & $2.07588 - 0.134619 \, i$ & $2.06643 - 0.135559 \, i$  \\ \hline
$4$    & $2.43315 - 0.135379 \, i$  & $2.44854 - 0.134684 \, i$ & $2.44221 - 0.138805 \, i$  \\ \hline
$5$    & $2.81279 - 0.134979 \, i$  & $2.82523 - 0.134556 \, i$ & $2.82053 - 0.143357 \, i$  \\ \hline
$6$    & $3.19382 - 0.134657 \, i$  & $3.20425 - 0.134375 \, i$ & $3.20389 - 0.143435 \, i$  \\ \hline
$7$    & $3.57575 - 0.134387 \, i$  & $3.58472 - 0.134186 \, i$ & $3.58429 - 0.144436 \, i$  \\ \hline
$8$    & $3.9583 - 0.134157 \, i$   & $3.96617 - 0.134008 \, i$ & $3.9698 - 0.145597 \, i$   \\ \hline
$9$    & $4.34129 - 0.13396 \, i$   & $4.34831 - 0.133845 \, i$ & $4.35221 - 0.144408 \, i$  \\ \hline
$10$   & $4.72462 - 0.133788 \, i$  & $4.73094 - 0.133697 \, i$ & $4.73747 - 0.146576 \, i$  \\ \hline
\end{tabular}
\caption{Prompt ring down modes for the first overtone number $n=0$ and $r_s=1$, $r_b=0.8$, $q=0.5$, $Q=2$, $p=\mu=0$ for various $\ell$.}\label{tabb6}
\end{table}
\begin{table}[H]
\centering
\begin{tabular}{|c|c|c|c|}
\hline
$r_b$ & $WKB$                  & Leaver               & $SW_4$                   \\ \hline
$0.5$ & $1.67435 - 0.180019 \, i$ & $1.72041 - 0.170672 \, i$ & $1.70821 - 0.156853 \, i$ \\ \hline
$0.6$ & $1.67723 - 0.166767 \, i$ & $1.71766 - 0.159428 \, i$ & $1.7079 - 0.144709 \, i$  \\ \hline
$0.7$ & $1.67991 - 0.152547 \, i$ & $1.71465 - 0.147257 \, i$ & $1.70806 - 0.130406 \, i$ \\ \hline
$0.8$ & $1.68218 - 0.13711 \, i$  & $1.71123 - 0.133916 \, i$ & $1.71388 - 0.108494 \, i$ \\ \hline
$0.9$ & $1.6836 - 0.120151 \, i$  & $1.70704 - 0.119066 \, i$ & $1.71739 - 0.121633 \, i$ \\ \hline
\end{tabular}
\caption{Prompt ring down modes for the first overtone number $n=0$ and $r_s=1$, $\ell=2$, $q=0.5$, $Q=2$, $p=\mu=0$ for various $r_b$.}\label{tabb7}
\end{table}

\subsection{Amplification factor}
The general solution at infinity \eqref{solinfpm} has to be matched with \eqref{sol1}, encoding ingoing boundary conditions at the horizon. This happens if 
\begin{align}
A_{+}{=}&\frac{\Gamma \left(\frac{1}{2}{-}m_3{-}\sqrt{u}\right) \Gamma \left(\frac{1}{2}{-}m_3{+}\sqrt{u}\right)}{ \Gamma
   \left(\frac{1}{2}{-}m_3{+}\sqrt{u}\right) \Gamma \left(\frac{1}{2} + m_3{-}\sqrt{u}\right){-}  ({-}1)^{2 \sqrt{u}}\Gamma \left(\frac{1}{2} {-} m_3{-}\sqrt{u}\right) \Gamma
   \left(\frac{1}{2} {+} m_3{+}\sqrt{u}\right)}\times\\\nn
   &\Bigg[\frac{ \Gamma \left(2 \sqrt{u}\right) \Gamma \left(\frac{1}{2}{-}m_1{-}\sqrt{u}\right) \Gamma \left(\frac{1}{2}{+}m_2{-}\sqrt{u}\right)
   \Gamma \left(\frac{1}{2}{+}m_3{-}\sqrt{u}\right)}{\Gamma \left({-}2 \sqrt{u}\right)}+\\\nn
   &{-}\frac{\Lambda
   ^{2 \sqrt{u}} \Gamma \left({-}2 \sqrt{u}\right)  \Gamma \left(\frac{1}{2}{-}m_1{+}\sqrt{u}\right) \Gamma \left(\frac{1}{2}{+}m_2{+}\sqrt{u}\right) \Gamma
   \left(\frac{1}{2}{+}m_3{+}\sqrt{u}\right)}{\Gamma \left(2 \sqrt{u}\right)}\Bigg]
\end{align}
\begin{align}
A_-{=}&\frac{\Gamma \left(\frac{1}{2}{+}m_3{-}\sqrt{u}\right) \Gamma \left(\frac{1}{2}{+}m_3{+}\sqrt{u}\right)}{ \Gamma
   \left(\frac{1}{2}{-}m_3{+}\sqrt{u}\right) \Gamma \left(\frac{1}{2}{+}m_3{-}\sqrt{u}\right){-} ({-}1)^{2 \sqrt{u}} \Gamma \left(\frac{1}{2}{-}m_3{-}\sqrt{u}\right) \Gamma
   \left(\frac{1}{2}{+}m_3{+}\sqrt{u}\right)}\times\\\nn
   &\Bigg[\frac{ \Gamma \left(-2 \sqrt{u}\right) \Lambda ^{2 \sqrt{u}} \Gamma \left(\frac{1}{2}{-}m_1+\sqrt{u}\right) \Gamma
   \left(\frac{1}{2} {+} m_2+\sqrt{u}\right) \Gamma \left(\frac{1}{2}{-}m_3{+}\sqrt{u}\right)}{\Gamma \left(2 \sqrt{u}\right)}+\\\nn
   &-\frac{\Gamma \left(2
   \sqrt{u}\right) \Gamma \left(\frac{1}{2}-m_1-\sqrt{u}\right) \Gamma \left(\frac{1}{2}+m_2-\sqrt{u}\right) \Gamma
   \left(\frac{1}{2}-m_3-\sqrt{u}\right)}{\Gamma \left(-2 \sqrt{u}\right)}\Bigg]
\end{align}
At infinity, the behaviour of \eqref{solinfpm} is
\be
\Psi(z) \sim Y_+  e^\frac{\Lambda z}{2} z^{m_3} {+} Y_- e^{-\frac{\Lambda z}{2}} z^{- m_3}
\ee

\be
Y_+ = - (-1)^{\frac{1}{2}+ m_3 - \sqrt{u}}\Lambda^{m_3} A_+ \quad,\quad Y_-= \Lambda^{-m_3}A_-
\ee
The amplification factor is defined as \cite{Brito:2015oca, Consoli:2022eey} 
\be
Z =\Bigl| \frac{Y_+}{Y_-}  \Bigr|^2 -1
\ee
Introducing the quantity 
\be \label{eq:Qgrav}
Q_{\rm{grav}} = \frac{qQ - \omega r_s}{\sqrt{1- \frac{r_b}{r_s}}} 
\ee
in the near-decoupling limit ($\L \ll 1$), the dictionary \eqref{dictBS} becomes 
\be\label{approxdic}
m_1 = - m_2 = -\text{i} Q_{\text{grav}}, \qquad m_3 = \frac{q}{2} \sqrt{r_s^2 - Q^2}, \qquad u = \left(\ell + \frac{1}{2}\right)^2 
\ee
Recalling we also work in the probe limit $q \simeq 0$, we impose $m_3 \rightarrow 0$.

In these approximations, the amplification factor can be written as:
\begin{equation}
   \label{Z-Sigma}
   Z \sim  \left|\frac{1- \Sigma}{1 + \Sigma}\right|^2-1 \sim -4 \text{Re}\Sigma 
\end{equation}
where
\begin{equation}\label{sigma0}
\Sigma=\frac{\Lambda^{2\sqrt{u}}\Gamma\left({1\over2}+\sqrt{u}\right)\Gamma\left({1\over 2}-m_1+\sqrt{u}\right)\Gamma\left({1\over2}+m_2+\sqrt{u}\right)\Gamma\left(-2\sqrt{u}\right)^2}{\Gamma\left({1\over2}-\sqrt{u}\right)\Gamma\left({1\over 2}-m_1-\sqrt{u}\right)\Gamma\left({1\over2}+m_2-\sqrt{u}\right)\Gamma\left(2\sqrt{u}\right)^2}
\end{equation}
Plugging in the previous equation \eqref{sigma0} the approximated dictionary \eqref{approxdic} and expanding for small $Q_{grav}$ we obtain:
\be
\Sigma=\frac{i(-1)^\ell 2^{-1-4\ell} Q_{grav}\Lambda^{2\ell+1}\Gamma\left(1+\ell\right)^4}{ \Gamma(2+2\ell)^2\Bigl[\left({\frac{1}{2}}\right)_\ell\Bigr]^2}
\ee
\begin{figure}
    \centering
    \includegraphics[width=0.8\linewidth]{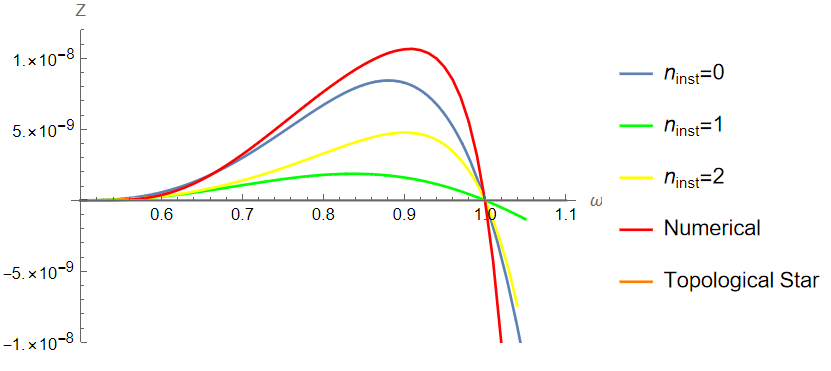}
    \caption{Comparison among different BS amplification factors: the red line represent the curve obtained by numerical integration while blue, yellow and green curves represent \eqref{Z-Sigma} with \eqref{amplA} and with 0, 1 and 2 instantonic corrections respectively. The chosen parameters are $r_s=1$, $r_b=0.8$, $\ell=2$, $q=0.5$, $Q=2$, $p=\mu=0$. The orange line refers to the amplification factor of TS with the same parameters as BS with $r_b$ and $r_s$ exchanged.}\label{plotAmpli}
  \end{figure}
In order to take the instantonic corrections into account we have to perform the replacement $\ell\rightarrow a-\frac{1}{2}$:
\be\label{amplA}
\Sigma=\frac{\pi ^2 (-1)^a 2^{1-8 a} Q_{grav} \Lambda ^{2 a} \Gamma
   \left(a+\frac{1}{2}\right)^2}{\Gamma (a)^2 \Gamma (a+1)^2}
\ee
Unfortunately, as it is shown in Figure \ref{plotAmpli}, the first instantonic corrections do not get closer to the numerical curve (see Appendix \ref{appA}), stating that the convergence would require a larger number of instantonic corrections. In the same Figure, we display the amplifaction factor for TS and it is clear that it is zero within the numerical precision of the software employed in the calculation (Wolfram Mathematica).

The superradiance here observed for the charged non-rotating BS is not linked to the superradiant instability highlighted in \cite{Press:1972zz, Cardoso:2005gj, Cardoso:2005vk}: indeed, the BS lacks the reflection mechanism needed to trap and enhance the outgoing perturbation.

\section{Conclusions and Outlook} \label{Conclusion}
Starting from the suggestion of \cite{Bianchi:2023sfs}, we have performed a detailed analysis of linear massive scalar \textit{charged} perturbations both for Topological Stars (TSs) and for Black Strings (BSs). The probe used in order to accomplish this goal is a stringy object winding around the compact dimension $y$.

The wave equation describing the dynamics of this problem is a CHE with two regular singularities (at the cap $r=r_b$ and at the horizon $r=r_s$) and one irregular singularity at infinity. Several methods have been implemented with the aim of computing QNMs. One of them consists in exploiting the correspondence with qSW curves for $\mathcal{N}=2$ SYM with $SU(2)$ gauge group and $N_f=(1,2)$ fundamental hypermultiplets and another one is the Leaver method. These two have been used both for TSs and BSs, while other techniques have been peculiarly employed for the two types of solutions.

For TSs, a direct numerical integration is used, as well. The QNMs determined with the three approaches are reported in Tables \ref{Tabb1}, \ref{Tabb2} and \ref{Tabb3}. Concerning BSs, the matching asymptotic expansion for near extremal and near superradiant modes is used, while for prompt ring down modes a WKB approximation is exploited. The near extremal and near superradiant QNMs computed with the previous three approaches are reported in Tables \ref{Tab 2 BS} and \ref{Tab 3 BS}, while the prompt ring down modes are presented in Tables \ref{tabb6} and \ref{tabb7}. We find good agreement among the applied methods. 

As it can be seen, for all the cases $\text{Im}\omega <0$, confirming the stability with respect to linearized \textit{charged} and scalar perturbations. Despite TS and JMaRT soliton are both non-BPS smooth horizonless geometries, the former seems not to be affected by charged linear instability. Then, as argued in \cite{Bianchi:2023rlt}, the culprit of JMaRT ergoregion and charge instability should not to be identified with non-extremality, but with the over-rotation. 

In addition, we studied the charge superradiance of TSs and BSs. 
As exposed in \cite{Brito:2015oca}, rotational superradiance occurs for geometries with a horizon inside an ergoregion. Our focus, instead, has been on static backgrounds which then can only give rise to a charge superradiance - very much as Reissner - Nordstr\"om \cite{Brito:2015oca, DiMenza:2014vpa, Benone:2015bst, Baake:2016oku} - due to the presence of a horizon and a charged probe with respect to the electromagnetic fluxes of the solutions\footnote{We are aware of the fact that superradiance can also take place for horizonless geometries \cite{Brito:2015oca}, leading to superradiant instabilities. Nonetheless, as highlighted in this paper, this does not happen for TSs.}. Our results confirm charge superradiance for BSs; TSs, instead, do not superradiate, as expected for fuzzball geometries which conserve information. This is reported in Figure \ref{plotAmpli}. The first instantonic corrections are taken into consideration for the computation of the amplification factor of BS. As it can be argued from the Figure \ref{plotAmpli}, we expect that the convergence is reached by increasing the number of instantonic corrections. This agrees with results shown in \cite{DiRusso:2024hmd}. 

We have also had the opportunity to study the Tidal Love Numbers of TSs and BSs in the case of charged probes. Focusing on the former, we find that these numbers exhibit the same peculiarities of the ones for the case of neutral probes as analysed in \cite{DiRusso:2024hmd}, that is there are poles in correspondence of metastable QNMs. 

We have established the stability of these geometries only for linear scalar perturbations. As a future development, we aim to investigate the stability of modes of higher spins (e.g. vector bosons and gravitons) or in the non-linear case. 

\section*{Acknowledgments}
We are grateful to M. Bianchi and J. F. Morales for their invaluable support through helpful discussions and insightful comments on the manuscript.  We also thank A. Argenzio, Y. F. Bautista, I. Bena, V. Bevilacqua, D. Bini, G. Bonelli, V. Collazuol, V. Cuozzo, S. De Angelis, A. De Santis, G. Dibitetto, G. Di Ubaldo, R. Dulac, A. Evangelista, D. Fioravanti, G. Frittoli, F. Fucito, C. Gambino, C. Gasbarra, A. Grassi, L. Grimaldi, C. Iossa, F. Margari, A. Ruiperez, L. Tabarroni, A. Tanzini, A. Tokareva for fruitful scientific exchanges. G.D.R. thanks IPhT Paris-Saclay for the kind hospitality during completion of this work.
\section*{Appendix}

\appendix
\section{Electric field of a charged wire} \label{Charged wire}
Consider an infinitely extended and uniformly charged wire along the $y$ direction in $\mathbb{R}^{1,D-1}$. The metric of this space is
\be \label{flatmetric}
\mathrm{d}s^2 = -\mathrm{d}t^2 + \mathrm{d}y^2 + \mathrm{d}r^2 + r^2 \mathrm{d}\Omega_{D-3}^2
\ee
where $\mathrm{d}\Omega_{D-3}^2$ is the metric of a $(D-3)$ - sphere of unitary radius. The transverse space $\mathcal{M}^{D-2}$ has the topology of $\mathbb{R}^+ \times S^{D-3}$. For a $(D-3)$-sphere embedded in a $(D-2)$-dimensional (flat) space, we introduce $D-3$ angles $\theta_1,\dots, \theta_{D-3}$ and only one of them is azimuthal ($\theta_{D-3} \in [0,2 \pi)$), while all the other ones are polar ($\theta_1,\dots, \theta_{D-4} \in [0,\pi)$). The induced metric is
\be
\label{induced metric}
\mathrm{d}\Omega^2_{D-3} = \sum_{j=1}^{D-3} \left(\prod_{i=1}^{j-1} \sin^2{\theta_i}\right) \mathrm{d}\theta_j^2
\ee 
In the transverse space the problem presents a radial symmetry, so the electric gauge potential sourced by the wire is
\be
B^{(e)} = -\Phi(r) \, \mathrm{d}t \wedge \mathrm{d}y
\ee
Hence, the electric field strength is
\be
F^{(e)} = \mathrm{d} B^{(e)} = E(r) \, \mathrm{d}r \wedge \mathrm{d}t \wedge \mathrm{d}y
\ee
The Hodge dual is
\be
\star F^{(e)} = - E(r) \, r^{D-3} \sqrt{|g_{S^{D-3}}|} \, \mathrm{d}\theta_1 \wedge \dots \wedge \mathrm{d}\theta_{D-3}
\ee
where $g_{S^{D-3}}$ is the determinant of \eqref{induced metric}. By Maxwell's equations\footnote{In this expression we have used that, for a generic $k$-form $A_k$ in a D-dimensional spacetime,
\be
\star \star A_k = (-)^s (-)^{k(D - k)} A_k
\ee
For \eqref{flatmetric}, $s=1$ and $\mathrm{d} \star F^{(e)}$ is a $(D-2)$-form.  
}, 
\be \label{Maxwell}
\mathrm{d}\star F^{(e)} = -\star j^{(e)}
\ee
and the 2-form electric current is $j^{(e)} = -\frac{\rho}{\epsilon_0} \mathrm{d}t \wedge \mathrm{d}y$, with $\rho$ the electric volume density, so 
\be
\star j^{(e)} = \frac{\rho}{\epsilon_0} \, r^{D-3} \, \sqrt{|g_{S^{D-3}}|} \, \mathrm{d}r \wedge \mathrm{d} \theta_1 \wedge \dots \wedge \mathrm{d} \theta_{D-3}
\ee
Due to the fact that $\rho$ is the electric volume density, then $\rho = \mathcal{Q}/(L_y V^{D-2})$, where $\mathcal{Q}$ is the electric charge along the length $L_y$ of the wire and $V^{D-2}$ is the volume of the transverse space. So $\rho = \lambda/V^{D-2}$, with $\lambda$ linear electric density, and consequently
\be
\label{d starFe}
\int_{\mathcal{M}^{D-2}} \mathrm{d} \star F^{(e)} = - \int_{\mathcal{M}^{D-2}}  \star j^{(e)} = - \frac{\lambda}{\epsilon_0 V^{D-2}} V^{D-2} = - \frac{\lambda}{\epsilon_0}
\ee
On the other side, thanks to Stokes' theorem, we can write
\be
\int_{\mathcal{M}^{D-2}} \mathrm{d} \star F^{(e)} = \int_{S^{D-3}} \star F^{(e)}
\ee
where we are assuming that the transverse space $\mathcal{M}^{D-2}$ has as boundary the ($D-3$)-sphere at fixed $r$ embedded in it.
So
\be
\label{starFe}
\int_{S^{D-3}} \star F^{(e)} = - E(r) r^{D-3} \int_{S^{D-3}} \sqrt{|g_{S^{D-3}}|} \, \mathrm{d}\theta_1 \wedge \dots \wedge \mathrm{d}\theta_{D-3} = - V(S^{D-3}) E(r) r^{D-3}
\ee
Equalling \eqref{d starFe} with \eqref{starFe}, we get
\be
E(r) = \frac{\lambda}{V(S^{D-3}) \epsilon_0 r^{D-3}}
\ee
It is known that
\be
V(S^{D-3}) = \frac{2 \pi^{\frac{D-2}{2}}}{\Gamma\left(\frac{D-2}{2}\right)} 
\ee
and so
\be
\label{electric field}
E(r) = \frac{\lambda \Gamma\left(\frac{D-2}{2}\right)}{2 \epsilon_0 \pi^{\frac{D-2}{2}} r^{D-3}}
\ee
Notice that for $D=4$ we get the well known result of standard Electromagnetism in $1+3$ dimensions.\\
Comparing \eqref{electric field} with $Q/r^2$ of \eqref{eq:top} \footnote{We stress that, even if the probe has finite length while now we have considered an infinite length of the wire, the quantity $F^{(e)}$ is dimensionally the same.}, we realize that the parameter $Q$ has the dimensions of a linear charge density and in measure units such that
\be
\frac{1}{4 \pi \epsilon_0} = 1
\ee
they are the same.

\section{Measure units}\label{appmeas}
We choose to work in measure units such that 
\be
c = G_5 = \frac{1}{4 \pi \varepsilon_0} = 1
\ee
In this framework, from Schwarzschild-Tangherlini metric \cite{Tangherlini:1963bw}, we gain 
$$[M] = [L]^{D-3} = [L]^2$$ 
Consequently, the force has dimension $[F]=[M] \, [L]^{-1} = [L]$ and, from (\ref{electric field}), $[F] = [q] [E] = [q]^2 [L]^{-3}$, we get that the dimension of the electric charge is 
$$[q] = [L]^2$$
Finally, the dimension of the linear charge density is
$$[Q] = [q] [L]^{-1} = [L]$$
The fact that $[M] = [L]^2$ and $[Q] = [L]$ is consistent with the conditions \eqref{eq:TopCharge} and \eqref{ADM mass}. Notice that in these measure units it is inconsistent to set $\hbar=1$ and indeed, due to the fact that $[E] = [M]= [L]^2$, then 
$$[\hbar] = [E] [L] = [L]^3$$

Let us notice that according to these measure units, the tension of a string has dimension 
\be
[T] = [M] [L]^{-1} = [L]
\ee
hence the constant $\alpha'$ is such that
\be
[\alpha'] = [T]^{-1} = [L]^{-1}
\ee
In this way, the action \eqref{eq:sigmamodel} has the correct dimension $[S] = [\hbar] = [L]^3$. Moreover the parameter $q$ defined in \eqref{eq:qdefinition} has dimension
\be
[q] = [L]^2
\ee
as expected for a charge and consistently with \eqref{eq:massshift}.

\section{Numerical amplification factor}\label{appA}
In this Appendix, we briefly describe the numerical procedure implemented in Mathematica used to plot the amplification curve. The starting point consists in finding the leading and a sufficient number of subleading behaviours at infinity and at the cap. At infinity, the radial wave function behaves as:
\begin{equation}\label{RINF}
    R_{\infty}(r)\simeq e^{i\epsilon \, r\sqrt{\omega^2-q^2}}r^{-1-\frac{i\epsilon(q^2r_s+2qQ\omega-(r_b+2r_s)\omega^2)}{2\sqrt{\omega^2-q^2}}}\sum_{n=0}^{N_{\infty}}d_n r^{-n}
\end{equation}
where $\epsilon=\pm1$ encodes both the behaviour at infinity and where we set $\mu=p=0$ for simplicity. 
The first coefficient is:
\begin{align}
    \frac{d_1}{d_0}=&\frac{q^2 \left(2 r_b+r_s\right)-\omega ^2 r_b-2 q Q \omega }{4 \left(q^2-\omega ^2\right)}-\frac{1}{8 \epsilon 
   \left(\omega ^2-q^2\right)^{3/2}}\Big[i (4 q^3 Q \omega  \left(2 r_b+3
   r_s\right)+\\\nn
   &-2 \omega ^2 \left(q^2 \left(3 r_b r_s+2 r_b^2+6 r_s^2\right)+2 \ell  (\ell +1)\right)-4 q Q \omega ^3 \left(r_b+2
   r_s\right)+\omega ^4 \left(4 r_b r_s+3 r_b^2+8 r_s^2\right)+\\\nn
   &+q^4 \left(3 r_s^2-4 Q^2\right)+4 q^2 \ell  (\ell +1))\Big]\\\nn
 &  \dots
\end{align}
For a TS $(r_b>r_s)$ we have to impose regularity at the cap, so the correct ansatz is:
\begin{equation}\label{RCAPts}
    R_c(r)\simeq \sum_{n=0}^{N_c}c_n(r-r_b)^n
\end{equation}
where
\begin{align}
    \frac{c_1}{c_0}=&-\frac{r_b(qQ-r_b\omega)^2-(r_b-r_s)\ell(\ell+1)}{(r_b-r_s)^2}\\\nn
    \frac{c_2}{c_0}=&\frac{1}{4(r_b-r_s)^4}\Big[3 \omega ^2 r_b^2 \left(2 q^2 Q^2 r_b^2+r_s
   \left(r_b-r_s\right)\right)-2 q Q \omega  r_b \left(r_b^2
   \left(2 q^2 Q^2+1\right)+r_b r_s-2 r_s^2\right)+\\\nn
   &{+}q^2
   \left(q^2 Q^4 r_b^2{+}Q^2 \left(r_b{-}r_s\right) \left(2
   r_b{+}r_s\right){+}r_b \left(r_b{-}r_s\right)^3\right){-}\ell^2
   \left(r_b{-}r_s\right) \left(r_b \left(2 \left(q Q{-}\omega 
   r_b\right)^2{+}1\right){-}r_s\right){+}\\\nn
   &{-}2 \ell 
   \left(r_b{-}r_s\right) \left(r_b \left(q Q{-}\omega 
   r_b\right)^2{+}r_b{-}r_s\right){-}4 q Q \omega ^3 r_b^5{+}\ell^4
   \left(r_b{-}r_s\right)^2{+}2 \ell ^3
   \left(r_b{-}r_s\right)^2{+}\omega ^4 r_b^6\Big]\\\nn
   &\dots
\end{align}
Otherwise for a BS the boundary conditions at the horizon require an ingoing wave:
\begin{equation}\label{RCAPbs}
    R_H(r)=(r-r_s)^{\frac{i(qQ-r_s\omega)}{\sqrt{1-\frac{r_b}{r_s}}}}\sum_{n=0}^{N_H}\tilde{c}_n(r-r_s)^n
\end{equation}
where 
\begin{align}
    \frac{\tilde{c}_1}{\tilde{c}_0}=&\frac{1}{\left(r_b{-}r_s\right)^2 \left(i \sqrt{1{-}\frac{r_b}{r_s}}{-}2 q Q{+}2 \omega  r_s\right)}\Big[\left(r_b{-}r_s\right) \left(\omega  r_s{-}q Q\right){+}i \sqrt{1{-}\frac{r_b}{r_s}} (q^2 \left(Q^2 r_b{+}r_s \left(r_b{-}r_s\right)^2\right){+}\\\nn
    &{+}2
   q Q \omega  r_s \left(r_s-2 r_b\right)+\omega ^2 r_s^2 \left(3 r_b-2 r_s\right)+\ell  (\ell +1)
   \left(r_s-r_b\right))\Big]\\\nn
  &\dots 
\end{align}

The numerical integration is performed starting from the horizon (or the cap) and proceeding up to infinity. If we denote $A_{\text{in}}$ and $A_{\text{out}}$ as the zero-order coefficients $d_0$ of the asymptotically expanded wave at infinity \eqref{RINF}, respectively corresponding to $\epsilon=-1$ and $\epsilon=1$, and if we compare this function with its numerical counterpart along with their derivatives, we obtain expressions for $A_{\text{in}}$ and $A_{\text{out}}$. The final step consists in plotting the quantity $Z=\left|\frac{A_{\text{out}}}{A_{\text{in}}}\right|^2-1$ as a function of $\omega$.

\FloatBarrier


\bibliographystyle{jhep}
\bibliography{bibINST.bib}

\end{document}